\documentclass[reprint,
superscriptaddress,
 amsmath, amssymb,aps,
floatfix,
]{revtex4-1}

\usepackage{graphicx}
\usepackage{dcolumn}
\usepackage{bm}
\usepackage[colorlinks=true,citecolor=blue,linkcolor=magenta]{hyperref}
\usepackage[utf8]{inputenc}
\usepackage{url}
\usepackage[normalem]{ulem}
\usepackage[mathlines]{lineno}
\usepackage{amsmath}

\begin{document}

\title{Pockels Soliton Microcomb}

\author{Alexander W. Bruch}
\affiliation{Department of Electrical Engineering, Yale University, New Haven, CT 06511, USA}

\author{Xianwen Liu}
\affiliation{Department of Electrical Engineering, Yale University, New Haven, CT 06511, USA}

\author{Zheng Gong}
\affiliation{Department of Electrical Engineering, Yale University, New Haven, CT 06511, USA}

\author{Joshua B. Surya}
\affiliation{Department of Electrical Engineering, Yale University, New Haven, CT 06511, USA}

\author{Ming Li}
\affiliation{Department of Optics and Optics Engineering, University of Science and Technology of China, Hefei, Anhui 230026, China}

\author{Chang-Ling Zou}
\affiliation{Department of Optics and Optics Engineering, University of Science and Technology of China, Hefei, Anhui 230026, China}

 \author{Hong X. Tang}
\affiliation{Department of Electrical Engineering, Yale University, New Haven, CT 06511, USA}
\affiliation{Corresponding author: hong.tang@yale.edu}


\begin{abstract}
Kerr soliton microcombs have recently emerged as a prominent topic in integrated photonics and enabled new horizons for optical frequency metrology. Kerr soliton microcombs, as its name suggests, are based on the high-order cubic optical nonlinearity. It is desirable to exploit quadratic photonic materials, namely Pockels materials, for soliton generation and on-chip implementation of $1f$-$2f$ comb self-referencing. Such quadratically-driven solitons have been theoretically proposed, but have not yet been observed in a nanophotonic platform despite of recent progresses in quadratic comb generation in free-space and crystalline resonators. Here we report photonic chip-based Pockels microcomb solitons driven by three-wave mixing in an aluminum nitride microring resonator. In contrast to typical Kerr solitons, our Pockels soliton features unity soliton generation fidelity, two-by-two annihilation of multi-soliton states, favorable tuning dynamics, and high pump-to-soliton conversion efficiency.
\end{abstract}

\maketitle
\section{Introduction}
Within the past decade, Kerr microcombs have made a phenomenal transition from laboratory curiosity to a field of study in its own right \cite{kippenberg2011microresonator}. Tight optical confinement and facile geometrical dispersion control afforded by nanophotonic waveguides have realized significant milestones in nonlinear optics and comb development in a chip-scale package \cite{gaeta2019photonic}. Today, these ``gold standard'' Kerr combs are reaching into application spaces such as optical frequency synthesis \cite{spencer2018optical}, optical clocks \cite{papp2014microresonator, newman2019architecture}, exoplanet observation \cite{suh2019searching}, chemical spectroscopy \cite{picque2019frequency, dutt2018chip}, and quantum optics \cite{kues2019quantum}. Many of these applications leverage the low-noise Kerr soliton state, where the comb lines become mutually phase-locked and form an optical pulse in the time domain \cite{herr2012universal, kippenberg2018dissipative, stern2018battery}. The exploitation of the lowest order $\chi^{(2)}$ optical nonlinearity can bring about additional novel phenomena in frequency comb development. Frequency doubling of a Kerr comb allows for efficient near-visible \cite{guo2018efficient, he2019self} and ultraviolet \cite{liu2019beyond} comb generation as well as direct detection of a comb's carrier envelope offset frequency \cite{hickstein2017ultrabroadband}. Nonlinear coupling between the fundamental and second harmonic bands also introduces novel comb dynamics not seen in the case of a typical Kerr comb \cite{guo2018efficient, xue2017second}. 

\begin{figure*}[!t]
    \centering
    \includegraphics[width=\textwidth]{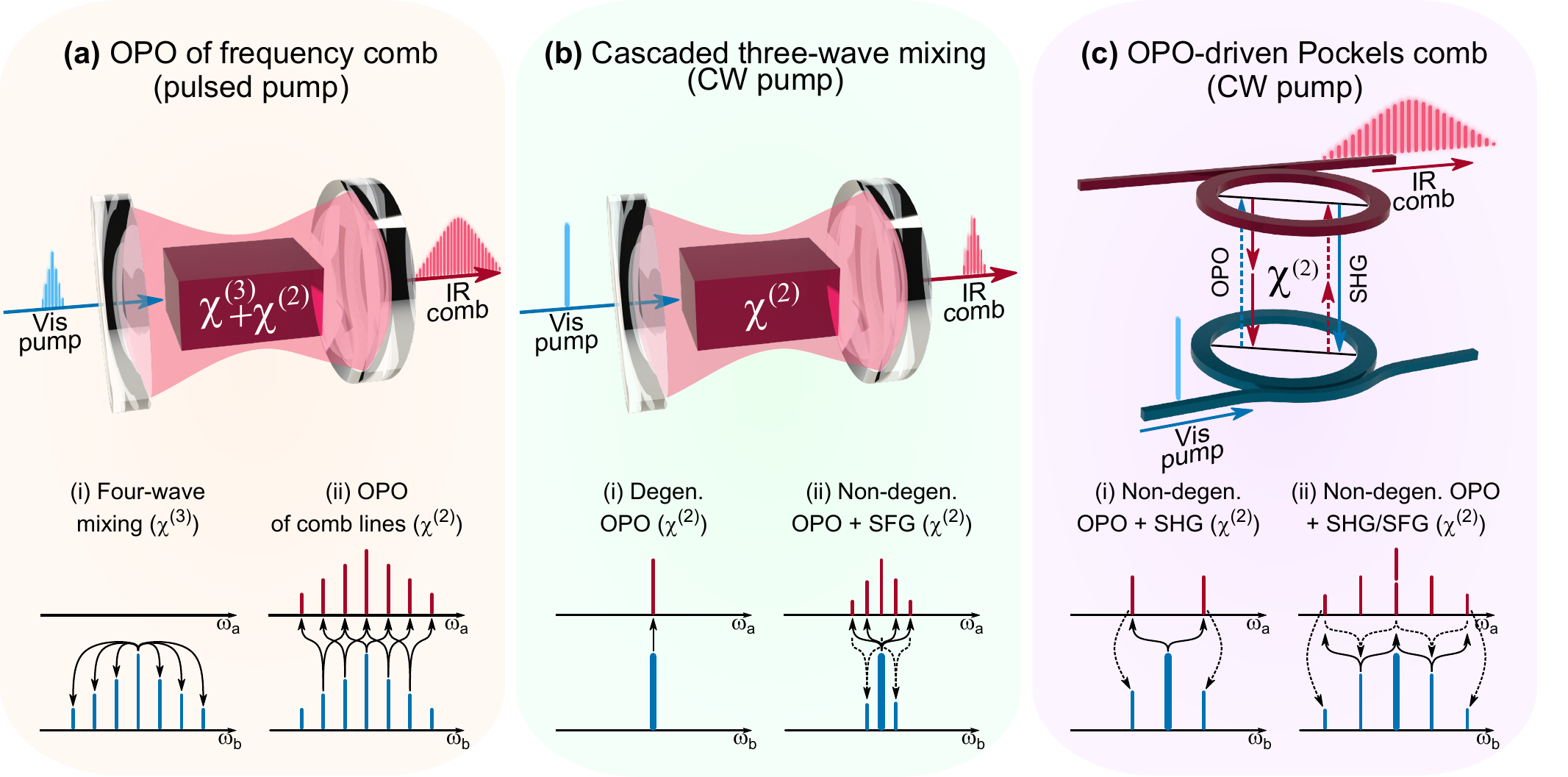}
	\caption{Approaches to $\chi^{(2)}$ OPO combs. (a) Broadband IR comb generation via OPO of a frequency comb in a cavity possessing both $\chi^{(2)}$ and $\chi^{(3)}$ nonlinearity. An ultrafast pulsed pump produces broadband spectra via four-wave mixing. A broadband comb is produced by OPO. (b) Narrowband IR comb generation from a CW pump via $\chi^{(2)}$ OPO in a doubly-resonant bulk cavity. Additional sidebands appear near the pump due to SFG between OPO lines. (c) Quadratic comb generation from a CW pump in a coupled microring system possessing strong $\chi^{(2)}$ nonlinearity. A near-visible pump produces IR frequencies via OPO similar to (b), while simultaneous SHG phase-matching produces near-visible frequencies. Back-to-back OPO, SHG, and SFG produce a frequency comb at both the pump and OPO frequencies. Schematics of the comb lines in the infrared ($\omega_a$, red) and near-visible bands ($\omega_b$, blue) are shown below each process in (i-ii).}
	\label{fig:comb_intro}
\end{figure*}

Frequency comb-like spectra arising solely from $\chi^{(2)}$ nonlinearity have been theoretically predicted for over two decades \cite{etrich1997solitary}, and were recently demonstrated in bulk optical cavities containing only $\chi^{(2)}$ nonlinearity
\cite{ulvila2013frequency, ricciardi2015frequency, Mosca2016, mosca2018modulation}. These combs arise from back-to-back second-harmonic/sum-frequency generation (SHG/SFG) and optical parametric oscillation (OPO). The process may also proceed in the reverse case (OPO followed by SHG/SFG) as in Refs.~\cite{ulvila2013frequency, mosca2018modulation}. Cascaded three-wave mixing between frequency components yields comb lines separated by the cavity free spectral range (FSR), as in the case of the Kerr frequency comb. These so-called ``quadratic frequency combs'' enjoy a relatively low power threshold and high pump-to-comb efficiency due to the inherently large magnitude of the $\chi^{(2)}$ susceptibility compared to the $\chi^{(3)}$ one. Additionally, the cascaded SHG/SFG-OPO process creates a frequency comb at both the fundamental and harmonic frequencies, naturally giving rise to dual-band comb spectra.

The potential benefits of quadratic frequency combs have ushered a flurry of numerical studies on the dynamics of combs driven by $\chi^{(2)}$ nonlinearity in micro-scale systems. Numerical simulations of quadratic combs have revealed modulation-instability (MI) regimes and nonlinear bistability akin to a Kerr frequency comb \cite{leo2016frequency, mosca2018modulation}, as well as low-noise Turing patterns including multiple soliton states \cite{hansson2018quadratic, villois2019soliton, villois2019frequency}. Experimental evidence of compact quadratic combs have only recently been observed in a 20~mm-long periodically-poled lithium niobate (PPLN) waveguide \cite{ikuta2018frequency} as well as MgO-doped lithium niobate bulk whispering gallery mode resonators \cite{hendry2019experimental, szabados2019frequency}. While prominent MI-like comb structures and a very low quadratic comb threshold was successfully observed in Ref.~\cite{szabados2019frequency}, quadratic solitons have not yet been experimentally observed.

Here, we report successful generation of a quadratic microcomb soliton in an aluminum nitride (AlN) nanophotonic platform.
To mitigate the competition with Kerr comb formation at the telecom band, we pump the device in the near-visible band near 780 nm and produce a stable quadratic solitons in the near-infrared regime near 1560 nm via OPO. 
Our numerical simulations describe a unique OPO detuning model that suggests the presence of multiple temporal patterns with distinct comb structures before the single-soliton state sets in. Additionally, our OPO-driven Pockels microcomb soliton features a deterministic soliton generation pathway and high pump-to-soliton conversion efficiency compared to its Kerr soliton counterpart. Our work suggests the great potential of Pockels soliton microcombs for future nonlinear photonic applications.

\section{Results}

Figure~\ref{fig:comb_intro} illustrates the various methods to access OPO frequencies in a resonant cavity system.  
The traditional route of producing frequency combs near the OPO frequency, particularly in the mid-infrared, uses a high-power, femtosecond pulsed pump \cite{schliesser2012mid, herr2018frequency}, as in Fig.~\ref{fig:comb_intro}(a). In a cavity containing both $\chi^{(3)}$ and $\chi^{(2)}$ nonlinearities, four-wave mixing expands the frequency comb through a Kerr ($\chi^{(3)}$) process  (Fig~\ref{fig:comb_intro}(a-i)), which is then spectrally translated to the OPO frequency through $\chi^{(2)}$ coupling (Fig.~\ref{fig:comb_intro}(a-ii)). OPO combs may also arise from a strong CW tone pumping a doubly-resonant cavity containing a $\chi^{(2)}$ medium (Fig.~\ref{fig:comb_intro}(b)). Cavity resonance near the degenerate OPO frequency excites sufficient nonlinear gain to produce emissions in the OPO band (Fig~\ref{fig:comb_intro}(b-i)). If the $\chi^{(2)}$ phase-matching bandwidth is sufficiently broad, both degenerate and non-degenerate OPO phase-matching may be satisfied simultaneously, resulting in a narrowband comb near the OPO frequency shown in Fig.~\ref{fig:comb_intro}(b-ii) \cite{boyd2003nonlinear}. Additional sidebands may arise near the pump due to SFG of the OPO comb lines.

Our on-chip implementation of an AlN microring system is idealized in Fig.~\ref{fig:comb_intro}(c), modeled as two mode families at near-visible (blue) and infrared (red) bands coupled by the $\chi^{(2)}$ effect, akin to a triply-resonant OPO \cite{guo2018efficient, bruch2019chip}. A Pockels microcomb builds upon the system in Fig.~\ref{fig:comb_intro}(b) when the $\chi^{(2)}$ nonlinearity is sufficiently strong to support simultaneous OPO and SHG, rather than SFG (Fig.~\ref{fig:comb_intro}(c)). As the power of the infrared signal and idler increases, side bands arise in the near-visible band via SHG of the signal and idler (Fig.~\ref{fig:comb_intro}(c-i)). These lines grow until reaching their respective OPO threshold, after which they begin mediating cascaded three-wave mixing that is requisite for Pockels comb generation. The infrared and near-visible combs expand via cascaded OPO and SHG/SFG, respectively, shown in Figure~\ref{fig:comb_intro}(c-ii). A background-free bright soliton may then arise when the OPO modes have anomalous dispersion \cite{villois2019frequency}.

\begin{figure*}[htb]
    \centering
    \includegraphics[width=\textwidth]{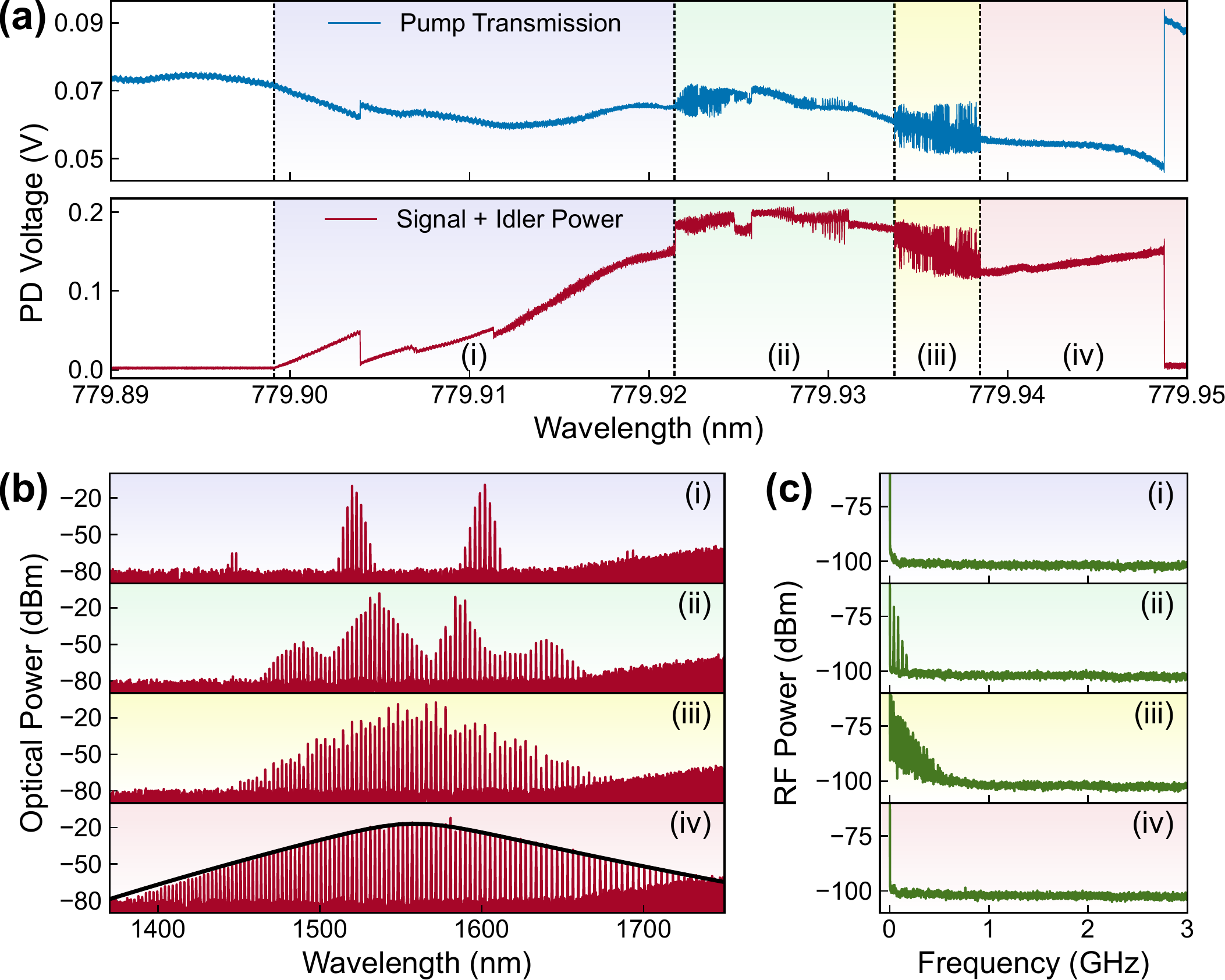}
	\caption{Comb detuning and power spectra. (a) Near-visible transmission (blue) and IR power (red) recorded as the microring resonance is scanned from blue-detuned to red-detuned states. Four distinct comb states are observed, and labeled (i-iv). Optical (b) and RF (c) spectra of the comb states observed in (a). A sech$^2$ fit is applied to the optical spectrum in (iv). The RF spectra in (i) and (iv) are equivalent to the measurement noise floor.}
	\label{fig:comb_detuning}
\end{figure*}

\noindent \textbf {Comb Characterization} 
Our experimental apparatus is sketched in Fig.~\ref{fig:comb_setup} in Methods, which consists of a dually-coupled AlN microring resonator that is phase-matched between 1560\,nm and 780\,nm modes. This device is pumped by a tunable Ti:Sapphire laser centered around 780~nm and exhibits a 12\,mW OPO threshold  \cite{bruch2019chip}. To investigate quadratic comb behavior, we set the temperature of an external heater beneath the chip for the degenerate phase-matching condition at low power and then boost the pump power to approximately 5 times the OPO threshold \cite{bruch2019chip}. Figure \ref{fig:comb_detuning}(a) shows the OPO power trace as the pump is scanned over the resonance at a moderate speed of $\thicksim$25 GHz/s. We observe Kerr comb-like distinct discontinuities and assign them into four different operation regimes labeled (i-iv), corresponding to the distinct comb states in Fig.~\ref{fig:comb_detuning} (b). We note that all four comb states exhibit relatively slow dynamics and can be accessed by manually tuning the laser frequency, in contrast to most Kerr comb generation processes which require tailored frequency tuning schemes on the order of THz/s to map various comb states \cite{Gong2018}.

State (i) lies just above the OPO threshold; its optical spectrum in Fig~\ref{fig:comb_detuning}(b-i) reveals features of non-degenerate OPO combs due to a cascaded OPO, sum-frequency generation (SFG), and difference frequency deneration (DFG) with the 780 nm pump similar to Fig.~\ref{fig:comb_intro}(b) in the non-degenerate case. The near-visible spectrum shown in Supplementary Note 1 shows little SHG of the OPO combs on the order of picowatts. We do not attribute the non-degenerate combs to four-wave mixing, as the infrared power ($\thicksim$12 mW) is below the estimated Kerr comb threshold ($\thicksim$40 mW). Furthermore, a numerical simulation excluding $\chi^{(3)}$ effects predicts similar non-degenerate comb behavior, although the presence of $\chi^{(3)}$ effects may expand the combs through non-degenerate four-wave mixing (See Supplemental Note 3). The jumps in the OPO power trace correspond to the progression of non-degenerate OPO combs towards degeneracy near 1560 nm. The non-degenerate OPO combs exhibit low RF noise near the noise floor of our measurement setup, shown in Figure~\ref{fig:comb_detuning}(c-i). This is expected because the FSRs of signal and idler combs are larger than the detector bandwidth and there is no spectral overlap between these clustered combs. 

The microcomb then transitions to state (ii), shown in Figure~\ref{fig:comb_detuning}(b-ii) and (c-ii). This comb is identified by the appearance of two strong non-degenerate OPO tones near 1520 nm and 1590 nm that merge near  degeneracy at 1560 nm. The RF spectrum of this comb exhibits a low overall noise floor with strong beat notes at 42 MHz and its harmonics. This beating suggests the comb spectrum in Fig.~\ref{fig:comb_detuning}(b-ii) consists of a superposition of two non-degenerate OPO combs centered at 1520 nm and 1590 nm with a slight frequency offset on the order of tens of MHz near 1560 nm. This offset is within the IR resonance linewidth of $\thicksim$200 MHz, which still fulfills the OPO phase-matching condition with a nonzero detuning from the cavity resonance frequency.  Beating of additional comb lines is not observed within the detector bandwidth. 

The non-degeneracy is pushed towards the degenerate OPO condition by further increasing the pump wavelength. The OPO comb power then enters a noisy, unstable state (iii) akin to modulation instability noise characteristic of a Kerr comb \cite{herr2012universal}. OSA and RF spectra in Fig.~\ref{fig:comb_detuning}(b-iii) and (c-iii) reveal a comb spectrum resembling Kerr MI comb with the maximum power centered at 1560 nm and pronounced RF noise in the low frequency regime. In the $\chi^{(2)}$-driven case, such MI-like noise is attributed to FSR locking between the $1f$ and $2f$ frequency components \cite{mosca2018modulation, szabados2019frequency}. Inspection of the dual-band comb spectra in Supplementary Note 1 reveals that both infrared and near-visible microcombs lock to an FSR of 361 GHz, limited by the resolution of the OSA. Upon further pump tuning, a low-noise soliton state (iv) is achieved after the noisy MI state. The optical power spectrum in Fig~\ref{fig:comb_detuning}(b-iv) reveals a characteristic sech$^2$ spectral envelope. The RF power spectra in Fig.~\ref{fig:comb_detuning}(c-iv) reveals RF noise near the measurement noise floor. Unlike a conventional Kerr soliton, the Pockels soliton optical spectrum is CW-background-free due to large spectral separation from the pump frequency. Applying a sech$^2$ fit to the soliton spectra reveals a 6.8 THz 3-dB optical bandwidth, corresponding to a 46 fs soliton pulse width. The spectral bandwidth of the soliton state is ultimately limited by the cavity dispersion at the OPO frequency ($D_2/2\pi \approx 15$ MHz), as predicted by previous theoretical studies \cite{villois2019soliton, villois2019frequency}. 

The spectra in Fig.~\ref{fig:comb_detuning}(b) demonstrate the progression of OPO phase-matching from non-degenerate (states i-ii) to degenerate OPO states (states iii-iv). Typically, degenerate and non-degenerate OPO comb states are accessed by varying the phase-matching temperature of the device \cite{mosca2018modulation}. We investigate the impact of the initial phase-matching condition by varying the chip temperature at a constant pump power in Supplementary Note 2. When tuned away from the degenerate phase-matching temperature, only non-degenerate comb states are supported and soltions such as that in Fig.~\ref{fig:comb_detuning}(b-iv) are not observed. While these non-degenerate states show low noise akin to states (i-ii), our current results suggest that quadratic solitons are only achieved when the degenerate OPO phase-matching is satisfied. Our numerical simulation in Supplemental Note 3 suggests that single soliton states are possible in the case of large phase mis-matching at relatively high pump powers, however these pump powers are difficult to achieve in the current experiment and lead to significant instabilities in the comb spectrum.


\noindent \textbf{Numerical Simulation} 
While the dynamics of the Kerr microcomb system can be modeled by the Lugiato-Lefever equation with significant accuracy \cite{herr2014temporal}, such an analytical solution does not yet exist for the Pockels comb system. A sech-pulse solution was recently proposed for the OPO-driven case \cite{villois2019frequency}, however the solution derived therein neglects the Kerr effect and GVD near the pump frequency and assumes normal dispersion near the OPO frequency instead of the anomolous dispersion presented here. We instead use a modal expansion method of coupled mode equations including both $\chi^{(2)}$ and $\chi^{(3)}$ effects to gain insights on the spectra observed in Fig.~\ref{fig:comb_detuning}. Details on the numerical simulation can be found in the Methods section as well as in Ref.~\cite{guo2018efficient}.

\begin{figure*}[htb]
    \centering
    \includegraphics[width=\textwidth]{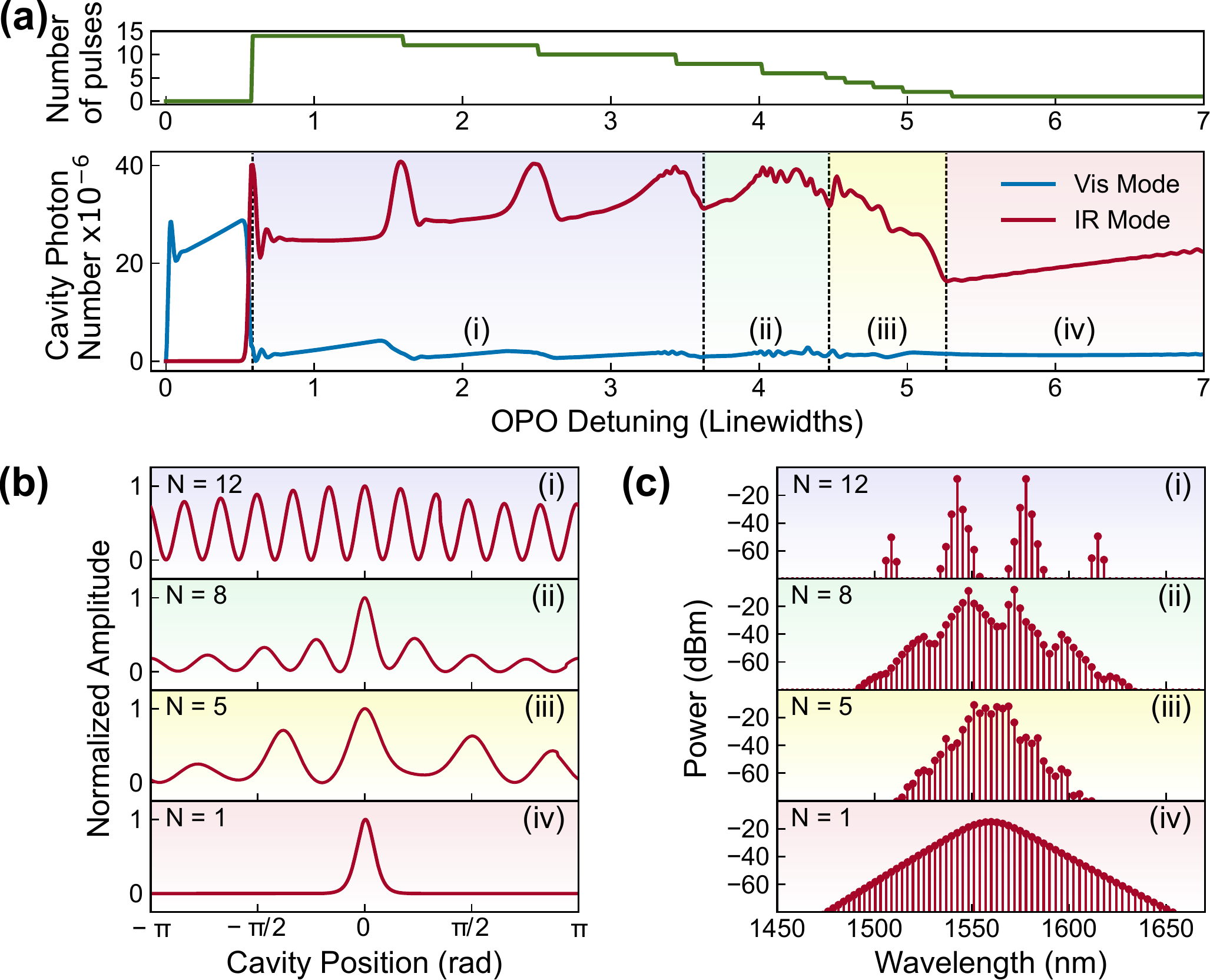}
	\caption{Numerical simulation of the Pockels microcomb. (a) Evolution of the near-visible (blue) and IR (red) cavity photon number as the pump is red-detuned. The number of optical pulses ($N$) is shown above as a reference. Comb states (i-iv) are labeled following the convention in Fig.~\ref{fig:comb_detuning}. Simulated temporal (b) and spectral (c) components of each comb state in (a). The evolution of the whole process can be viewed in the Supplementary Movie.}
	\label{fig:comb_simulation}
\end{figure*}
The experiment in Fig.~\ref{fig:comb_detuning} suggests that scanning the pump laser frequency changes the OPO phase-matching condition and pushes the Pockels microcomb into the degenerate OPO state. Given the large thermal shift of the cavity resonance in Fig.~\ref{fig:comb_detuning}(a) and that state (i) is always non-degenerate OPO, we assume the OPO process begins with an initial phase mis-match from the degenerate OPO condition of $\omega_b = 2 \omega_a$, where $\omega_{b(a)}$ is the frequency of the visible (infrared) mode. We indicate the detuning of the pump from the degenerate phase-matching condition as the "OPO detuning," described by $\omega_b - 2\omega_a = \xi \kappa_a$, where $\xi$ is the initial phase mis-matching parameter and $\kappa_a$ is the total amplitude decay rate of the infrared mode $a$. We then sweep the pump frequency $\omega_p$ throughout the simulation to observe the impact due to this OPO detuning. We note that this initial phase mis-match is distinct from the OPO phase-matching. The former is related to the initial detuning of the pump laser from degeneracy, wheras the latter describes the frequencies generated by the OPO process.

An exemplary numerical simulation of our Pockels cavity system under a strong visible pump is shown in Fig.~\ref{fig:comb_simulation}(a) for an initial phase mismatching of  $\xi = 14$. As the pump is scanned across resonance, the onset of OPO is observed as a sharp increase in infrared photon number with a corresponding drop in the near-visible photon number. Sharp changes in the infrared photon number are observed as the OPO detuning further decreases, corresponding to increasing degeneracy of the OPO-phase matching, similar to that observed in Fig.~\ref{fig:comb_detuning}(a). The time domain waveform of state (i) reveals a Turing pattern corresponding to the superposition of $N$ optical pulses in the microcavity. Fig.~\ref{fig:comb_simulation}(b-i) shows the temporal envolope for a non-degenerate OPO comb with $N=12$. The corresponding comb clusters in Fig.~\ref{fig:comb_simulation}(c-i) are separated by $N \cdot FSR$ from the degenerate OPO wavelength of 1560 nm. As the pump is further red-detuned, the number of pulses decreases 2-by-2 as $(N-2)$, $(N-4)$, $\cdots$; the spectral separation of each comb cluster from the degenerate OPO condition correspondingly decreases as $(N-2) \cdot FSR$, $(N-4) \cdot FSR$, $\cdots$. The reduction in pulse number is shown in the upper panel in Fig.~\ref{fig:comb_simulation}(a).

In the case of a typical Kerr comb, a Turing pattern of period $N$ corresponds frequency comb with $N \cdot FSR$ line spacing \cite{hu2015spatiotemporal} centered about the pump. Correspondingly, the period of a Turing pattern in a Pockels microcomb corresponds to the amount of detuning between the signal/idler from from the central degeneracy condition. Even-ordered Turing patterns are observed in the Pockels microcomb case due to the symmetric phase-matching condition of the signal and idler relative to the down-converted pump. 
The Pockels microcomb system is unique in that the degeneracy, and therefore Turing pattern period, changes with OPO detuning and dynamically reduces the pulse number throughout the comb tuning process. The evolution of the Turing patterns can be seen in real-time in the Supplemental Movie.

This process continues as the comb enters state (ii) as in Fig.~\ref{fig:comb_simulation}(c-ii). Even-numbered Turing patterns are still visible ($N=8$), and the pulse number continues to decrease in even orders with further OPO detuning. However, one pulse begins to grow in intensity as the comb lines close to degeneracy point become stronger, shown in Fig.~\ref{fig:comb_simulation}(b-ii). This strong pulse grows further in intensity as the comb enters the degenerate state (iii) and odd pulse numbers arise in the cavity starting at $N=5$ as shown in Fig.~\ref{fig:comb_simulation}(b-iii) and (c-iii). Here, the intracavity photon number and temporal amplitude exhibit considerable instability compared to states (i-ii) and the pulse number begins to decrease 1-by-1 until $N=1$, realizing a single pulse in the time domain. The corresponding infrared comb exhibits a sech$^2$ power spectrum similar to that shown in Fig.~\ref{fig:comb_detuning}(b-iv). This rapid re-organization of the pulse number likely gives rise to the MI-like noise observed in the experiment.  The OPO detuning process can be seen in its entirety in the Supplemental Movie. The near-visible temporal and spectral profiles in Figure~\ref{fig:comb_simulation}(b-c) may be found in Supplementary Note 3.

These results suggest that the Pockels comb process is driven primarily by $\chi^{(2)}$ processes in the microring. However, the small mode volume and anomalous dispersion in our AlN microring may allow $\chi^{(3)}$ effects to also play a role. The interplay between $\chi^{(2)}$ and $\chi^{(3)}$ effects in the OPO-driven Pockels comb has been studied in detail Ref.~\cite{villois2019frequency} as well as in Supplementary Note 3 by removing all $\chi^{(3)}$ interactions and re-running the simulation. Low-noise Turing patterns are observed in ``pure'' Pockels system, however single soliton states are not stable in the absence of $\chi^{(3)}$ effects. Moreover, the absence of $\chi^{(3)}$ effects yields a narrower comb profile at both infrared and near-visible bands compared to the case with both $\chi^{(2)}$ and $\chi^{(3)}$ effects. These findings suggest that the Pockels comb process is primarily driven by $\chi^{(2)}$ effects, while weak $\chi^{(3)}$ interactions further expand and stabilize the spectral and temporal profiles, respectively.

Our simulations indicate that the beating of OPO signal/idler-pairs in our system forms a regular periodic potential; the deterministic reduction in the pulse number in the 2-by-2 fashion described above smoothly induces soliton states without going through large MI noise. This smooth reduction in optical pulse number yields a nearly unity soliton generation fidelity with relatively slow dynamics as observed in Figure~\ref{fig:comb_detuning}. 

These unique dynamics exhibit traits that are distinct from the Kerr soliton. Primarily, the Pockels comb originates as a temporal pattern which reduces in pulse number before reaching the single soliton state. This is unlike the Kerr comb, which traverses through a CW-like MI state before reaching temporal pulses. The OPO detuning in a Pockels comb smoothly induces soliton states with a simple blue-to-red cavity scan, whereas a Kerr comb requires precise frequency ramping to avoid the chaotic MI state \cite{jaramillo2015deterministic}. Second, a relatively small drop in infrared comb power is observed in Fig.~\ref{fig:comb_detuning}(a) and Fig.~\ref{fig:comb_simulation}(a) as the comb transitions from the noisy state (iii) to the soliton state compared to that typically seen in a Kerr soliton. The process outlined above smoothly decreases the number of optical pulses in the microring, resulting in a small power drop as the pulse number decreases to the single soliton state. Intuitively, this is similar to a multi-soliton Kerr comb that exhibits relatively small drops in power as the comb transitions to the single soliton  \cite{kippenberg2018dissipative}. The Pockels comb system is unique in that it directly seeds multi-soliton states, avoiding the power loss seen in the Kerr comb system as the comb transitions from the MI to multi-soliton state. 

\begin{figure*}[t]
    \centering
    \includegraphics[width=\textwidth]{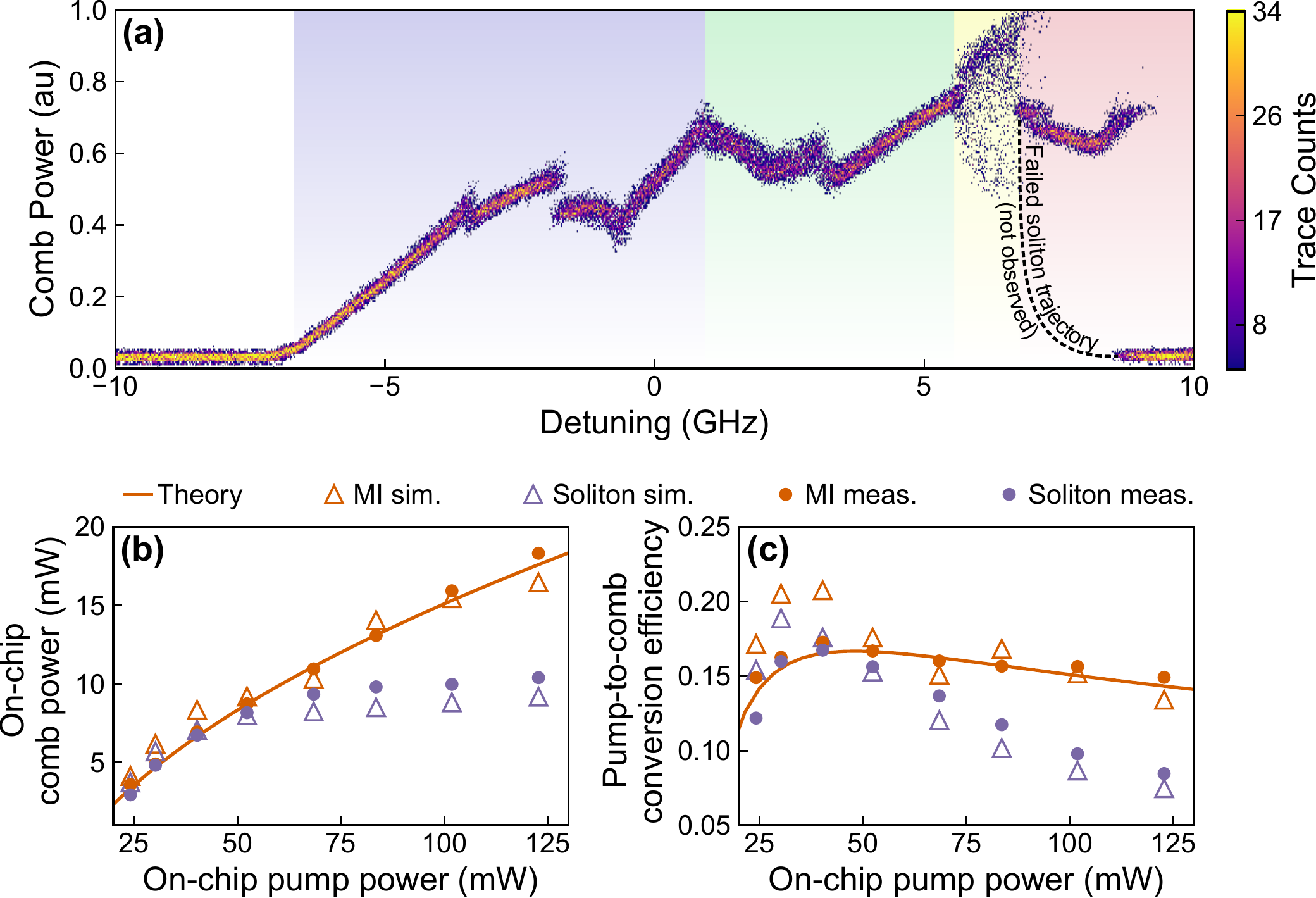}
	\caption{Robustness and power conversion efficiency of Pockels comb. (a) Histogram of 34 successive comb power traces. The color bar corresponds to probability of trace overlap. The trajectory of a failed soliton state is shown as a reference, which did not happen in our experiment.  On-chip OPO comb power (b) and pump-to-comb conversion efficiency (c). The MI comb power (orange) increases with near-visible pump power, while the on-chip soliton power (purple) saturates to 10 mW. The power and efficiency calculated by the numerical simulation is shown as orange and purple triangles for the MI and soliton states, respectively.}
	\label{fig:comb_benefits}
\end{figure*}

\begin{figure}[htb]
    \centering
    \includegraphics[width=\linewidth]{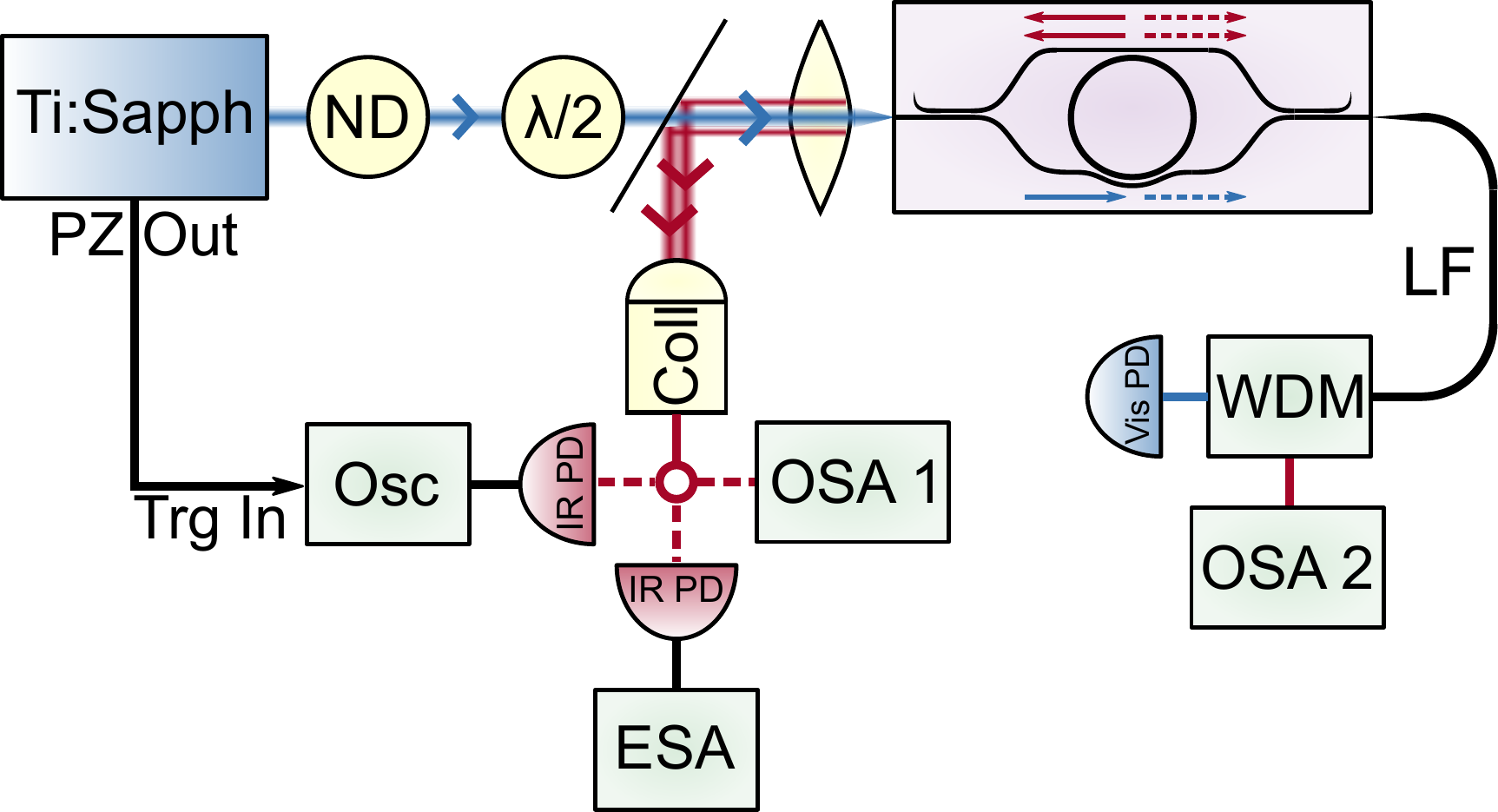}
	\caption{Simplified measurement schematic. Blue and red refer to near-visible and infrared light, respectively. Transparent lines refer to free-space beams. Dashed lines refer to optional measurements following the infrared collimator. Scattered infrared light is monitored on OSA 2 throughout the measurement. Descriptions of each component are given in the main text.}
	\label{fig:comb_setup}
\end{figure}

\noindent \textbf{Soliton stability and output power}
We further investigate the soliton stability and pump-to-comb power conversion ratio to provide an experimental comparison to our simulations above. Fig.~\ref{fig:comb_benefits}(a) displays an overlay of 34 successive comb power traces recorded at an on-chip power of $~$80 mW and a tuning speed of 25 GHz/s by scanning the laser cavity piezo.
The four OPO comb states in Fig.~\ref{fig:comb_detuning}(a) are observed for all 34 comb traces. A slightly reduced overlay density is observed in states (ii-iii), due to unstable RF and intensity noise as observed in Fig~\ref{fig:comb_detuning} above.
All of the traces then converge to the single soliton state.
We note that the high soliton fidelity was achieved using a simple cavity piezo scan, allowing for deterministic soliton generation without the need for fast pump tuning or pump power modulation schemes \cite{Gong2018, herr2012universal, brasch2016bringing, brasch2016photonic}.


The pump power dependence was then probed by varying the power coupled to the waveguide via a neutral-density filter. By controlling the phase-matching temperature, Pockels microcombs and solitons may be generated at a relatively low threshold of 25 mW on-chip near-visible power, albeit with reduced bandwidth compared to that in Fig.~\ref{fig:comb_detuning}(b). 
We note that the 25 mW quadratic comb threshold observed here is nearly half the estimated Kerr comb threshold at infrared wavelengths ($\thicksim$40 mW) and more than an order of magnitude lower than that at near-visible wavelengths ($\thicksim$370 mW) calculated using the optical $Q$ factors and $n_2$ of AlN at each wavelength \cite{liu2018integrated}. This low threshold is due to the relatively large magnitude of $\chi^{(2)}$ susceptibility compared to $\chi^{(3)}$; indeed the cascaded $\chi^{(2)}$ process has shown to be dominant over the $\chi^{(3)}$ at moderate optical \emph{Q} factors near $10^5$ \cite{boyd2003nonlinear}. Quadratic comb thresholds on the order of hundreds of $\mathrm{\mu}$W  have been reported in whispering gallery mode resonators \cite{szabados2019frequency}, comparable to the current state-of-the-art in Kerr microcomb systems \cite{Zhang2019}.

The absence of a large EDFA background commonly seen in Kerr microcombs allows us to directly collect the Pockels comb power near the OPO frequency. As the pump power further increases, the MI comb power increases up to 19 mW, following the characteristic OPO power curve \cite{Breunig2016, bruch2019chip} (orange line in Figs.~\ref{fig:comb_benefits}(b-c)). The soliton power begins to saturate to 10 mW after the pump power exceeds 75 mW.  We then numerically verify this behavior by sweeping the visible pump strength. The IR photon number continuously increases with power in the $N>1$ comb states, whereas the photon number begins to saturate in the $N=1$ soliton state when the pump power exceeds 60 mW. This power dependence is explicitly shown in Supplemental Note 3. The IR photon number is then converted to total IR power and largely agrees with the pump power dependence observed in the experiment, shown as the hollow triangle symbols in Figs.~\ref{fig:comb_benefits}(b-c). Our experimental and numerical results suggest the soliton power is clamped by an external mechanism unrelated to the traditional OPO process. At this time we believe the saturation of the soliton power is largely due to the dispersion of the sech$^2$ envelope of our system similar to that observed in the Kerr soliton case \cite{bao2014nonlinear}.
%

Normalizing the on-chip IR comb power to the on-chip near-visible power yields the on-chip efficiency shown in Fig.~\ref{fig:comb_benefits}(c). We observe a maximum efficiency of 17~\% for both the MI state and soliton state which then saturates to 15~\% in the MI state. The pump-to-soliton conversion efficiency rapidly depletes after the maximum value of 17~\% due to the saturated soliton power. Fig.~\ref{fig:comb_simulation}(a) highlights that the pump is largely converted to the OPO frequency above threshold, yielding the large pump-to-comb conversion efficiencies here. Furthermore, the OPO power and efficiency in states (i-iii) scale nearly exactly with that predicted by OPO theory \cite{Breunig2016, bruch2019chip}; intuitively, this can be thought of spreading the available OPO power across many frequency components instead of the typical signal and idler pair. This can be confirmed by comparing the OPO power in Fig.~\ref{fig:comb_detuning}(a) and the photon number in Fig.~\ref{fig:comb_simulation}(a) across states (i-iv).


\section{Discussion}

The broad interest in microcomb solitons merits a discussion on the key differences between the Pockels- and Kerr-driven combs. In summary, our results suggest the quadratic soliton excels where the Kerr soliton lacks. 
Our current results demonstrate a 17~\% conversion efficiency for Pockels solitons, while Kerr bright solitons often have $<$5~\% pump-to-soliton conversion efficiency under typical coupling conditions \cite{bao2014nonlinear}. 
The Pockels soliton also exhibits relatively slow dynamics, which can be probed by hand tuning of the pump wavelength. Our numerical simulations suggest multiple Turing pattern states throughout the OPO detuning process. The smooth reduction in optical pulses into the single soliton state results in a nearly deterministic soliton generation pathway. This process is dominated by $\chi^{(2)}$ phase-matching which can be well-controlled compared to the fast dynamics experienced in the case of a Kerr soliton which often require complex pump detuning and power ramping techniques to lock to the soliton state \cite{Gong2018}.

Our current modeling suggests that the degeneracy of the Pockels comb is controlled by initial phase mis-matching from the degenerate OPO condition. The OPO phase-matching condition changes with the laser frequency tuning and ultimately pushes the Pockels comb from the non-degenerate to the target degenerate phase-matching condition. As indicated by the ``flat bottom'' thermal triangle in Fig.~\ref{fig:comb_detuning}(a), the OPO initiates in state (i) very early in the thermal triangle and persists for a large range of frequency detunings. At this time, we believe the existence of the thermal triangle allows the OPO process to initiate with a large phase mis-matching compared to traditional OPO, giving rise to the detuning-dependent dynamics observed here. 

While thermal effects may also affect the OPO degeneracy \cite{bruch2019chip}, the device temperature is optimized for degenerate OPO phase-matching in the low-power regime; thermally-dominated phase-matching would push the device towards the non-degenerate phase-matching condition as the device temperature increases. The opposite effect is observed here, suggesting thermal effects do not play a significant role throughout the experiment. Further studies are needed to thoroughly investigate the interplay between thermal phase-matching and OPO detuning effects.

Despite the distinct benefits of the Pockels comb, the main advantage of the Kerr soliton is its large optical bandwidth. $\chi^{(3)}$ phase-matching may be readily optimized over a wide range of frequencies, realizing octave-spanning solitons in a chip-integrated platform \cite{pfeiffer2017octave, li2017stably}. Our current quadratic soliton has a relatively limited 3-dB bandwidth optical bandwidth of 6.8 THz. Similar to the Kerr case, the Pockels microcomb bandwidth is ultimately limited by second-order dispersion at the OPO frequency \cite{villois2019soliton, villois2019frequency}. Adoption of $\chi^{(2)}$ phase-matched geometries with optimal dispersion could realize broadband quadratic solitons. While simultaneous quadratic and cubic phase matching is difficult to realize by engineering the modal phase matching and group-velocity dispersion, it could be achieved in quasi-phase matched PPLN \cite{lu2019periodically, chen2019ultra} microring resonators 
for sustaining the quadratic comb over a broad frequency range. Octave-spanning quadratic combs would naturally open routes to facile $f-2f$ self referencing \cite{hickstein2017ultrabroadband}. The dual-band nature of quadratic solitons and high conversion efficiency would allow for direct detection of the comb $f_{ceo}$ if octave-spanning comb spectra could be achieved. Such quadratic combs would simplify comb stabilization for optical clocks \cite{papp2014microresonator, newman2019architecture} and exoplanet observation \cite{suh2019searching}.

\section{Methods}
\noindent \textbf{Measurement schemes}
We probe the whole process using the experimental setup shown in Fig.~\ref{fig:comb_setup}(b). The pump source is a free-space CW Ti:Sapphire laser ($700-1000$ nm, M2 SolsTiS), while the incident power and polarization are controlled via a neutral-density filter (ND) and half-wave plate ($\lambda/2$), respectively. The IR comb is mainly emitted towards the pump facet, which is then reflected off of a dichroic mirror and collected in a variable-zoom collimator (Thorlabs CFC-11X-C) for efficient fiber collection. The infrared comb is probed via a high-sensitivity optical spectrum analyzer (OSA 1, Yokogowa 6374) or collected in a high speed photodetector. The photodetector signal can then be viewed on an electrical spectrum analyzer (ESA, Advantest U3751), or in time-domain with an oscilloscope (Tektronix MDO3014). The Ti:Sapphire piezo tuning voltage (PZ Out) is used as a trigger for the time-domain measurement (Trg In). Approximately 1-5\% of infrared light is scattered towards the transmission port of the chip, which is collected via a lensed fiber and separated from the near-visible transmission via a fiber-based wavelength division multiplexer (WDM). The scattered infrared light is monitored on a separate OSA (OSA 2, Hewlett Packard 70004A) to record the comb spectra throughout the measurement. The near-visible transmission is monitored on a high-speed near-visible photodetector. 

\noindent \textbf{Numerical simulations}
The dynamics of the optical modes in the resonator are described by the Hamiltonian
\begin{equation}
\begin{split}
    \mathcal{H} = \sum_{j=-N_1}^{N_1} \hbar \Delta_j^a a_j^\dagger a_j + \sum_{j = -N_2}^{N_2} \hbar \Delta_j^b b_j^\dagger b_j \\ + \mathcal{H}_{\chi^{(2)}} + \mathcal{H}_{\chi^{(3)}} + \hbar \epsilon_p(b_0 + b_0^\dagger)
\end{split}
\end{equation}
where $a_j$ and $b_j$ represent the infrared and visible bosonic operators of the $j$th mode. An external near-visible pump with power $P_{in}$ and detuning from the cold-cavity resonance $\delta$ is introduced near mode $b_0$ with strength $\epsilon_p = \sqrt{\frac{2 \kappa_{b0,ex} P_{in}}{\hbar \omega_p}}$. The mode frequencies of each family are given by $\omega_j = \omega_0 + d_1j + \frac{d_2}{2} j^2$ and $\Omega_j = \Omega_0 + D_1j + \frac{D_2}{2} j^2$ for modes $a$ and $b$, respectively. $d_N$ and $D_N$ refer to their $N$th order dispersion coefficients. The detunings of each mode family relative to a pump field near $b_0$ with magnitude $\delta$ are given by $\Delta_j^a = \omega_0 - \Omega_0/2 +  \frac{d_2}{2} j^2 - \delta/2$ and $\Delta_j^b = (D_1 - d_1)j + \frac{D_2}{2} j^2 - \delta$. $\mathcal{H}_{\chi^{(2)}}$ denotes the $\chi^{(2)}$ effect in the microresonator with strength 
\begin{equation}
\mathcal{H}_{\chi^{(2)}} = \sum_{j,k,l} \hbar g_j^{(2)} (a_j a_k b_l^\dagger + a_j^\dagger a_k^\dagger b_l). 
\end{equation}
The sum $j + k = l$ denotes the phase-matching condition between modes $a_j$, $a_k$, and $b_l$. $\mathcal{H}_{\chi^{(3)}}$ represents the Kerr (within one mode family) and cross-Kerr (between mode families) effects. We consider the phase mis-matching due to the large thermal shift at high pump power via $\omega_b - 2\omega_a = \xi \kappa_a$, where $\xi$ is the initial phase mis-matching from the degenerate OPO condition $\omega_b = 2\omega_a$.

Under conditions of a strong external pump near mode $b_0$, the cavity field near the pump mode can be approximated as a classical field. 
The dynamics of the system are solved through the Heisenberg equation
\begin{equation}
\begin{split}
\frac{d}{dt}a_j = & (-i \Delta_j^a - \kappa_j^a)a_j - i\sum_{k,l} 2g_{jkl}^{(2)} a_k^\dagger b_l \\ 
& - i \sum_{kln}2g_{jkln}^{(3)aa} a_k^\dagger a_l a_n  - i \sum_{kln} g_{jkln}^{(3)ab} a_k b_n^\dagger b_l
\end{split}
\end{equation}
\begin{equation}
\begin{split}
\frac{d}{dt}b_j = & (-i \Delta_j^b - \kappa_j^b)b_j - i\sum_{k,l} 2g_{klj}^{(2)}a_k a_l \\   
& - i \sum_{kln}2g_{jkln}^{(3)bb}b_k^\dagger b_l b_n - i \sum_{kln} g_{nkjl}^{(3)ab}a_n^\dagger a_k b_l \\
& - i\epsilon_p \delta_D(b_0-p)
\end{split}
\end{equation}
where $\delta_D(b_0-p)$ is the dirac delta function representing detuning of the pump $p$ from cold-cavity resonance $b_0$. The Heisenberg equations are solved by the Fast Fourier Transform method to increase computational efficiency \cite{guo2018efficient} and are evolved in time by a fourth-order Runge-Kutta method. Time steps and detunings are normalized to $\kappa_a$. We take the dispersion values as $d_1/2\pi$ = 363 GHz, $D_1/2\pi$ = 351 GHz, $d_2/2\pi$ = 20 MHz, $D_2/2\pi$ = 34 MHz and neglect high order dispersion. The nonlinear coupling strengths are taken as $g^{(2)}/2\pi$ = 0.1 MHz, $g^{(3)}_{aa}/2\pi$ = 5.6 Hz, $g^{(3)}_{bb}/2\pi$ = 15 Hz, $g^{(3)}_{ab}/2\pi$ = 17 Hz. The nonlinear coupling rates and loss rates are assumed to be wavelength-independent due the our limited comb bandwidth. 




\section{Data availability} The data that support the findings of this study are available from the corresponding authors upon reasonable request.

\bibliography{soliton_refs}




\vspace{1 mm}
\noindent \textbf{Acknowledgements.} This work is supported by DARPA SCOUT (W31P4Q-15-1-0006). H.X.T. acknowledges support from DARPA's ACES programs as part of the Draper-NIST collaboration (HR0011-16-C-0118) and a NSF EFRI grant (EFMA-1640959) and David and Lucile Packard Foundation. The authors thank Yong Sun, Sean Reinhart, Kelly Woods, and Dr. Michael Rooks for assistance in the device fabrication. 

\vspace{1 mm}
\noindent \textbf{Author contributions.} A.W.B and H.X.T conceived the experiment design. A.W.B performed the device fabrication, measurement and data analysis with the assistance from X.L, Z.G, and J.B.S. M.L and C.Z performed numerical simulations and provided theoretical support. A.W.B and H.X.T. wrote the manuscript with the input from all other authors. H.X.T supervised the project.

\vspace{2 mm}
\noindent \textbf{Competing interests.} The authors declare no competing interests.


\clearpage

\onecolumngrid

\begin{center}

\Large

\vspace*{4cm}

\textbf{Supplementary Information \\ for ''Pockels Microcomb Soliton", Bruch et al.}

\end{center}


\maketitle

\setcounter{figure}{0} 
\renewcommand{\thefigure}{\textbf{\arabic{figure}}}
\renewcommand{\figurename}{\textbf{Supplementary Figure}}

\large
\textbf{Supplementary Note 1: Infrared and visible spectra of Pockels microcomb states}
\normalsize

\begin{figure*}[!h]
\centering
    \includegraphics[width = 0.9\textwidth]{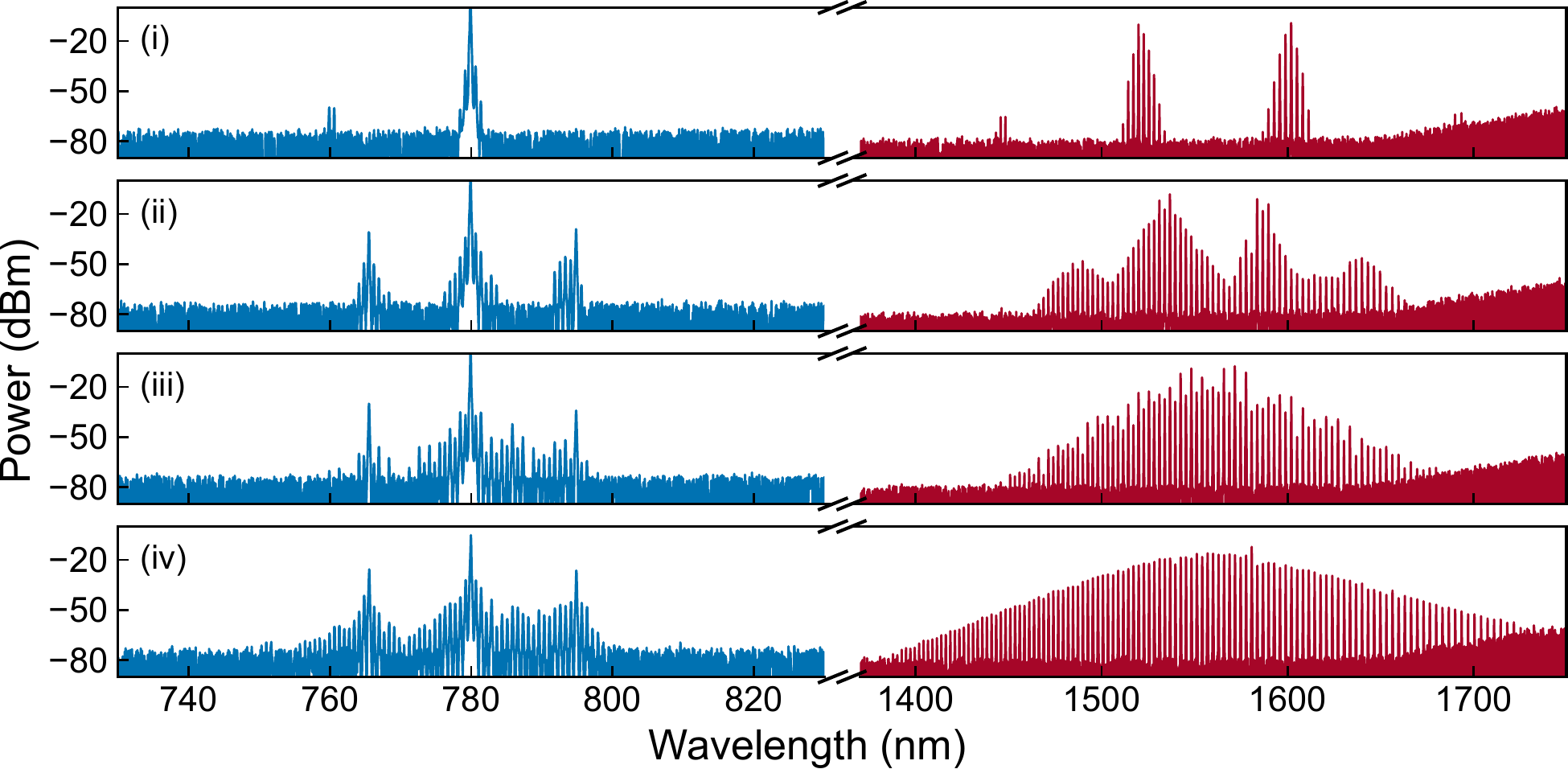}
    \caption{Dual band comb spectra. Near-visible (blue) and  Infrared (red) states corresponding to states (i-iv) shown in Fig.~2 in the main text.}
    \label{fig:combstates}
\end{figure*}

The spectrum of the visible comb collected from the transmission port of the AlN chip of the main text was also measured on OSA 2. The spectrum is measured after the WDM and measured using a double-monochromator on the OSA to minimize crosstalk between the infrared and visible spectra. Infrared spectra of the comb states in Fig.~2(b) in the main text are shown alongside their visible spectra in Fig.~S\ref{fig:combstates}.

State (i) corresponds to cascaded non-degenerate OPO as in Fig.~1(a) in the main text. Two non-degenerate OPO combs are produced from the visible pump, while sidebands arise near the 780 nm pump due to SFG between the infrared OPO comb frequencies. Very weak SHG of the OPO comb lines near 1520 nm is observed near 760 nm. As the comb transitions to the partially-degenerate OPO comb in state (ii), SHG of the non-degenerate OPO combs becomes apparent near 765 nm and 795 nm. Additional visible lines in the visible spectrum appear via SFG between the various OPO comb lines. The OPO comb then transitions to the degenerate MI state (iii). Interestingly, the near-visible lines stay pinned to their locations arising in state (ii) while simultaneously satisfying degenerate phase-matching with the 780 nm pump. In this degenerate state, SFG lines become more phase-matched near the pump frequency, filling in the gaps between the pump and the strong lines near 765 nm and 795 nm. Finally, SFG phase-matching across the entire visible comb bandwidth is achieved as the Pockels comb transitions to the soliton state (iv). The visible comb has a noticeably smoother profile than state (iii) due to the sech$^2$ IR soliton spectral profile, however its overall profile appears to be dominated by phase-matching rather than dispersion similar to the case of Cherenkov radiation observed in Ref.~\cite{guo2018efficient}.

\large
\textbf{Supplementary Note 2: Tunable degeneracy of the Pockels microcomb}
\normalsize

\begin{figure*}[!h]
\centering
    \includegraphics[width = 0.9\textwidth]{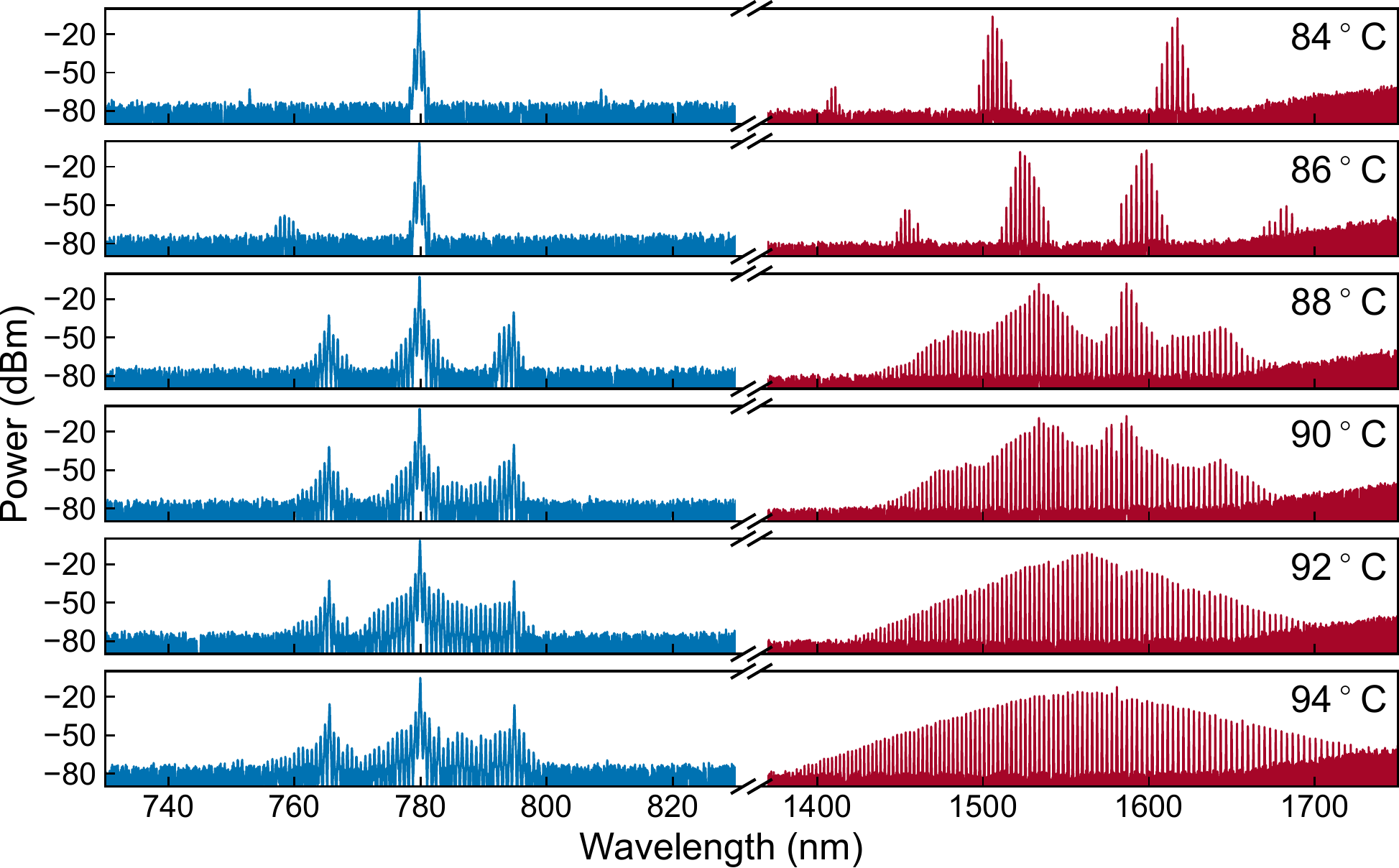}
    \caption{Temperature tuning of OPO comb spectra. Near-visible (blue) and infrared and (red) as the temperature is tuned from the degenerate OPO condition (bottom) towards the non-degenerate condition (top).}
    \label{fig:temp_tuning}
\end{figure*}

The OPO process has a widely tunable degeneracy that is often adjusted via temperature \cite{bruch2019chip}. We investigate the effect of degeneracy by probing the various comb states at a constant $\sim$75 mW on-chip pump power. Fig~S\ref{fig:temp_tuning}(a-b) shows the infrared and near-visible comb spectra as the temperature is tuned from its degenerate condition near 94$^\circ$C to the non-degenerate case near 84$^\circ$C. In general, the comb spectra follow the characteristic temperature dependence  observed in an OPO above threshold \cite{bruch2019chip}; a similar behavior was observed in a bulk cavity Pockels comb operating in the OPO mode under a CW pump \cite{mosca2018modulation}. Pockels solitons are observed in the degenerate condition near 94$^\circ$C with a nearly full visible comb. At 92$^\circ$C, the comb is pushed towards non-degeneracy and the the visible comb is less full near 770 nm. Only partially degenerate Pockels combs are supported as the temperature is further decreased to 90$^\circ$C and 88$^\circ$C. The visible comb spectra between the pump and the lines near 760 nm and 795 nm are noticeably less full and completely disappear at 88$^\circ$C. The infrared comb has a slightly wider optical bandwidth in this state with noticeable wings near 1470 nm and 1660 nm, at the sacrifice of less optical power near degeneracy at 1560 nm. The Pockels comb becomes fully non-degenerate at 86$^\circ$C and 84$^\circ$C with very little power phase-matched to the near-visible band.

The observed behavior allows us to gain some insights into the Pockels microcomb process. First, there appears to be a required SHG phase-matching bandwidth to sustain a broadband partially-degenerate or fully degenerate OPO comb. Our AlN microresonators have exhibited a broad OPO phase-matching bandwidth of over 180 nm \cite{bruch2019chip}, however the SHG phase-matching bandwidth is typically only $\sim$40 nm \cite{bruch2018}. Pockels combs are only possible when SHG provides sufficient power to re-oscillate to the OPO frequency and seed the cascaded $\chi^{(2)}$ process. This puts a bandwidth requirement on the Pockels comb phase-matching as $\Delta \omega_{OPO} \approx \Delta \omega_{SHG}$ such that the OPO phase-matching bandwidth ($2\Delta \omega_{OPO}$) falls within the SHG phase-matching bandwidth ($\Delta \omega_{SHG}$). Second, $\chi^{(2)}$ solitons are only possible when degenerate OPO phase-matching is satisfied. Our numerical simulations suggest that Pockels solitons emerge as the optical pulse number deterministically decreases from $(N  > 1)$ to $(N = 1)$. In the case of non-degenerate OPO phase-matching, the $(N = 1)$ cannot be satisfied, sustaining only the $N$ through $(N-M)$ number of optical pulses. This effect is studied in more detail in our numerical simulatin in Supplemental Note 3.

\large
\textbf{Supplementary Note 3: Details on the Numerical Simulation}
\normalsize

\begin{figure*}[!h]
    \includegraphics[width = 0.9\textwidth]{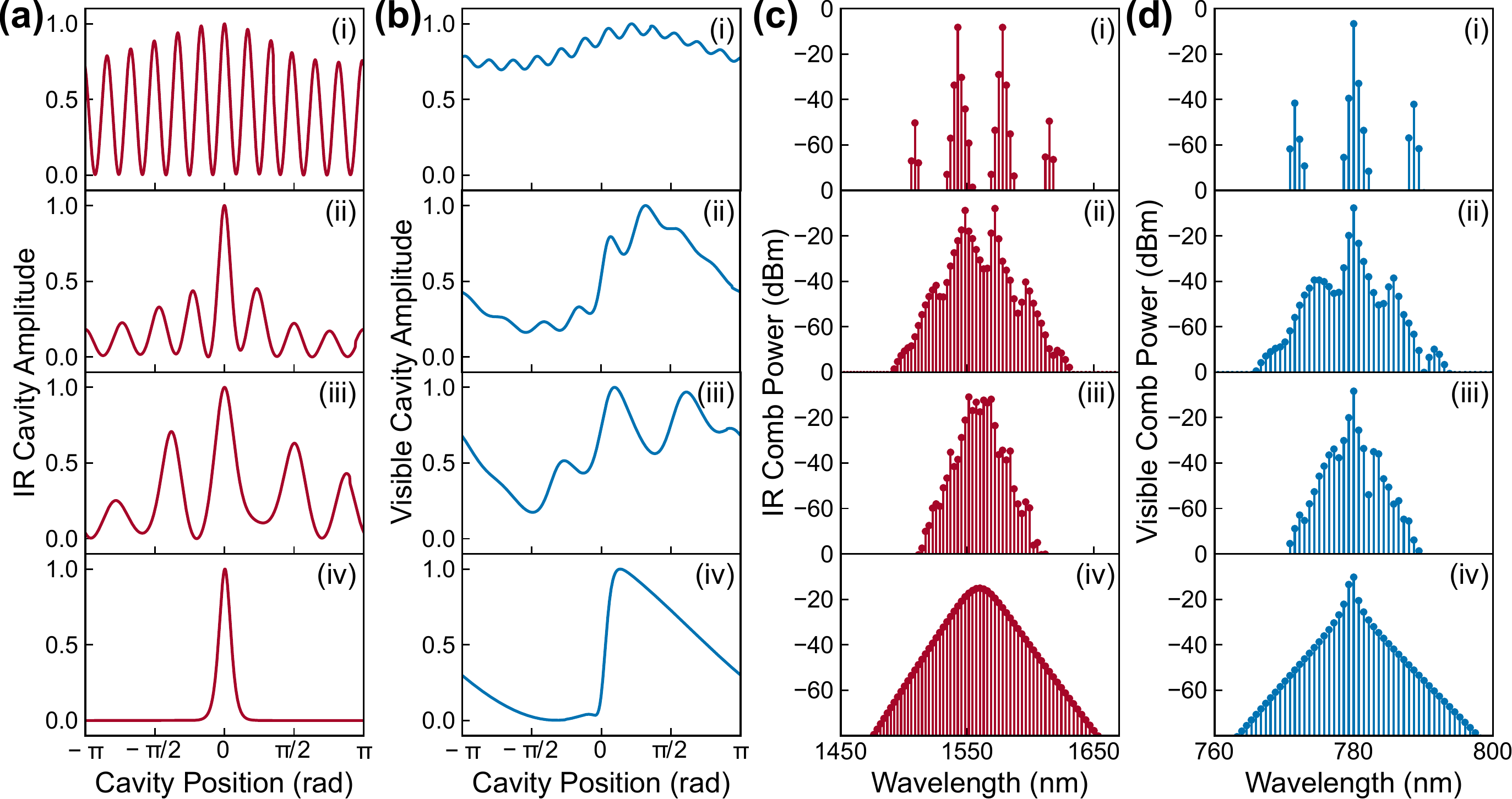}
    \caption{Simulated Infrared and and Visible Spectra of states i-iv in Fig.~3(b-c) of the main text. (a) Infrared and (b) near-visible temporal cavity amplitude. (c) Infrared and (d) near-visible spectral comb profiles.}
    \label{fig:sim_everything}
\end{figure*}

The near-visible temporal and spectral characteristics were also collected as in Fig.~3(b-c) of the main text. These spectra are shown side-by-side with the corresponding IR spectra for states (i-iv) in Fig.~S\ref{fig:sim_everything}. Upon entering state (i), both the infrared and near-visible temporal profile show a Turing pattern of $N=12$, corresponding to the non-degeneracy of the infrared comb in Fig.~S\ref{fig:sim_everything}(c-i). A large CW background is observed in the near-visible temporal profile due to the strong CW pump in this band. The corresponding near-visible comb spectrum shows SHG of the non-degenerate OPO combs near 770 nm and 790 nm, as well as SFG of the non-degenerate OPO comb near 780 nm. The near-visible temporal profile continues to show $N=8$ and $N=5$ pulses superimposed on a CW background corresponding to the infrared Turing pattern in states (ii-iii). The near-visible comb spectra are nearly degenerate in these states due to cascaded SHG and SFG of the OPO combs. Finally, the near-visible temporal profile shows an S-like temporal profile upon entering the soliton state (iv). While bright solitons are expected by the near-visible dispersion ($D_2/2\pi \approx$30 MHz), interference between the soliton and the CW pump purturbs the typical sech pulse profile, yielding the observed temporal profile. The corresponding visible comb spectrum shows a typical sech$^2$ comb profile for a bright soliton, as expected from the cavity dispersion.

The simulated near-visible comb spectra agree relatively well with those in Fig.~S\ref{fig:combstates}(b), matching SFG lines near the pump as well as the SHG of the non-degenerate OPO combs in Fig.~S\ref{fig:sim_everything}(d-i,ii). Notably, the simulated near-visible comb spectrum does not exhibit the strong peaks near 765 nm and 795 nm in Fig.~S\ref{fig:combstates}(b-ii,iii,iv). Nevertheless, the current simulation captures the SHG and SFG processes that drive the Pockels soliton process to arrive at the correct infrared temporal and spectral profiles. Our simulation does not currently take wavelength-dependent phase matching and/or wavelength-dependent coupling into account, both of which could give rise to the experimentally observed peaks in the near-visible band. Future iterations of our modeling would include such parameters to shed light on this phenomena.

\begin{figure*}[!h]
    \includegraphics[width = 0.9\textwidth]{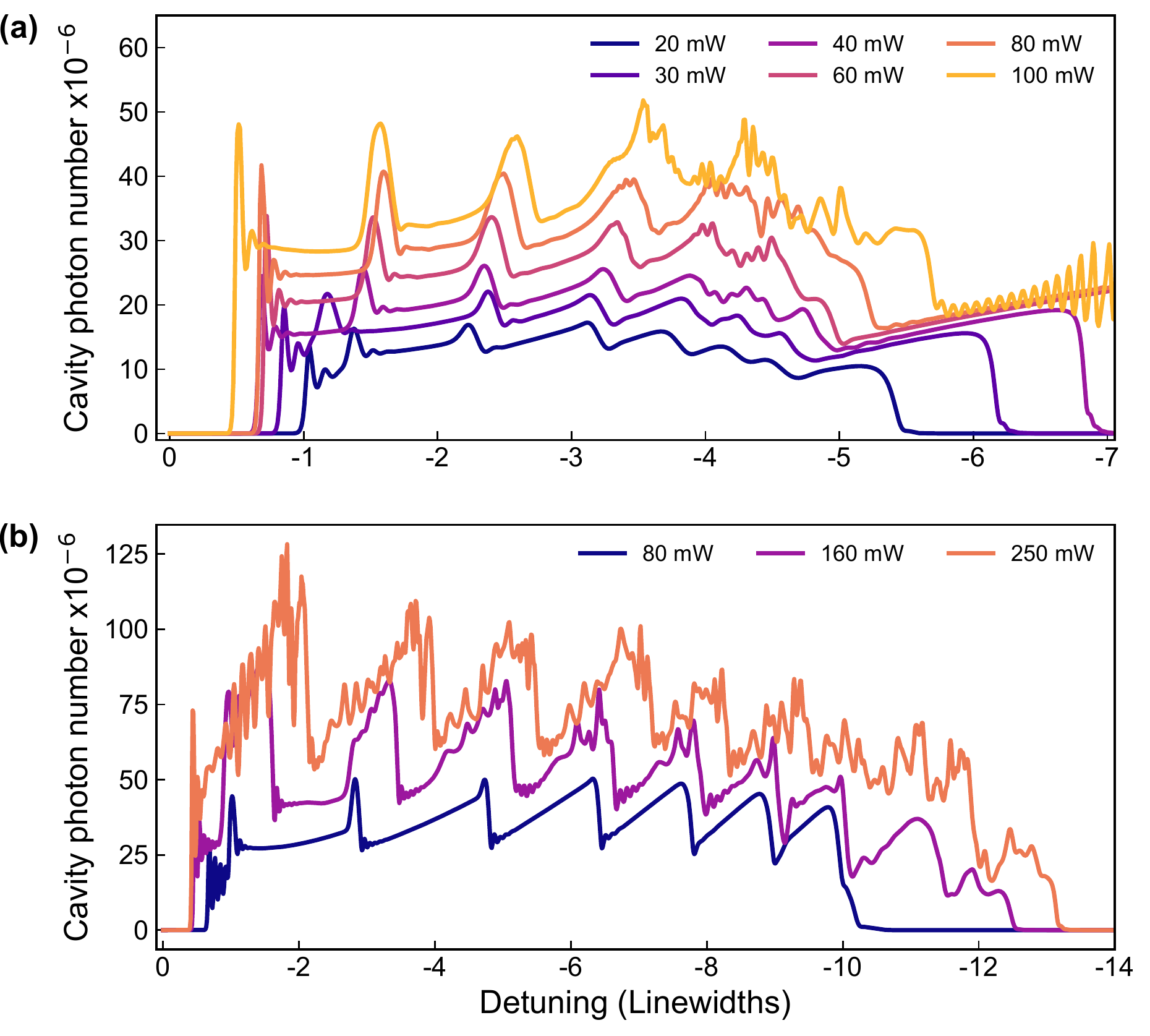}
    \caption{(a) Power dependence of the infrared cavity photon number for $\xi$=14. The soliton state is indicated by the step-like feature near -5 to -6 linewidths. (b) Infrared intracavity photon number for large phase mis-matching ($\xi$=50). Soliton states are indicated by the step-like features near -12 and -13 linewidths.}
    \label{fig:sim_powerdep}
\end{figure*}

The numerical simulation in Fig.~3 of the main text as well as Fig.~S\ref{fig:sim_everything} was performed under various pump strengths to obtain the simulated Pockels microcomb power in Fig.~4 of the main text. The intracavity photon number of the infrared mode is displayed in Fig~S\ref{fig:sim_powerdep}(a) to highlight the soliton power under various pump strengths. The data in Fig.~S\ref{fig:sim_powerdep}(a) is converted to power via $P_a = \hbar \omega_a \sqrt{\kappa_{a,1}} |a|^2$. The intracavity photon number increases with power in states (i-iii), following the OPO power relation (orange line line in Fig.~3(b-c) of the main text). The photon number begins to saturate to $\sim$20 x10$^6$ in the soliton state when the visible pump power exceeds 60 mW and limits the OPO power. This saturation effect has been theoretically predicted in the case of a Kerr comb and is limited by the dispersion of the sech$^2$ envelope \cite{bao2014nonlinear}. Additionally, both the MI and soliton states (states iii-iv) show significant instability at high pump power beyond 100 mW. These fluctuations may arise from competition between $\chi^{(2)}$ and $\chi^{(3)}$ effects that occurs at high pump powers \cite{villois2019frequency}.

The temperature dependence of the OPO degeneracy in Fig~S\ref{fig:temp_tuning} suggests that soliton states are only possible when the Pockels microcomb satisfies the degenerate OPO phase-matching condition. We simulate the effect of large phase mis-matching from degeneracy by increasing $\xi$ from 14 to 50 as in Figure~S\ref{fig:sim_powerdep}(b). At moderate pump powers of 80 mW, the Pockels comb tunes through many non-degenerate OPO comb states, shown by the sharp spikes in the cavity photon number. The comb power sharply decreases before reaching the single soliton state, mirroring the effect observed in Fig.~S\ref{fig:temp_tuning} above. Increasing the pump power to 160 mW and 250 mW reveals single soliton states near -12 and -13 linewidths, however this high power adds significant instability similar to the case of 100 mW in Fig.~S\ref{fig:sim_powerdep}(a). In practice, such high input powers would perturb the OPO phase-matching condition and may be difficult to realize experimentally.

\begin{figure*}[htb]
    \centering
    \includegraphics[width = 0.9\textwidth]{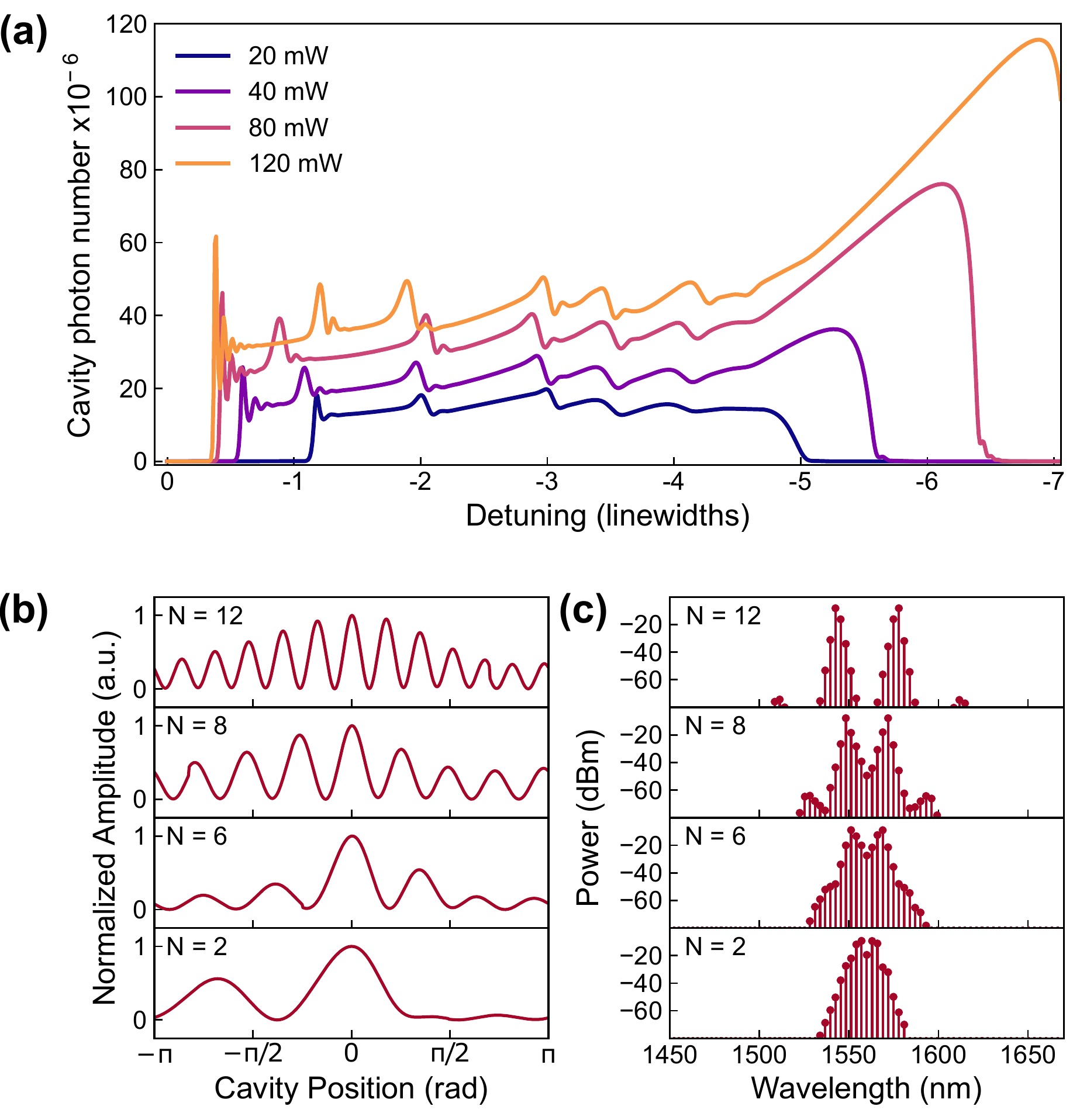}
    \caption{(a) Intracavity photon number for a quadratic microcomb without $\chi^{(3)}$ effects. The on-chip pump power is swept from 20 mW to 120 mW. (b) Temporal and spectral (c) profiles of the Pockels comb at an input power of 80 mW.}
    \label{fig:sim_nochi3}
\end{figure*}

Finally, we also simulate the case of a pure quadratic comb without $\chi^{(3)}$ effects by setting $g^{(3)}_{aa} = g^{(3)}_{bb} = g^{(3)}_{ab}= 0$ and running the simulation as in Figure~3 of the main text. The resulting intracavtiy photon number in Fig.~S\ref{fig:sim_nochi3}(a) shows similar discontinuities corresponding to the changing degeneracy of the non-degenerate OPO combs and a very pronounced soliton-like step. Interestingly, instabilities such as those in Fig.~S\ref{fig:sim_powerdep}(a) are not observed at a high pump power of 120 mW. This low-noise behavior suggests that four-wave mixing may play a role in the Pockels microcomb process when the pump and/or OPO power becomes significantly large. 

The temporal and spectral profiles are shown in Figure~S\ref{fig:sim_nochi3}(b) and (c), respectively. States (i-iii) are observed and have similar temporal and spectral profiles as the case with $\chi^{(3)}$ effects in Figure~3(b-c) of the main text. Despite the soliton-like step in Fig.~S\ref{fig:sim_nochi3}(a) the corresponding temporal profiles in Fig.~S\ref{fig:sim_nochi3}(b) show a $N=2$ turing pattern rather than a single soliton state and a much reduced comb bandwidth compared to Figures~3(c) in the main text and Fig.~S\ref{fig:sim_everything}(c-d). While $N=1$ soliton states have been theoretically proposed in a microtoroid system without $\chi^{(3)}$ effects \cite{villois2019frequency}, these are not captured by the current simulation. However, our numerical simulation does suggest that the $\chi^{(3)}$ effect does help induce single soliton states with a narrower temporal profile and correspondingly broader comb span compared to the pure $\chi^{(2)}$ case.

\end{document}